# High Helicity Vortex Conversion in a Rubidium Vapor

Aurélien Chopinaud, Marion Jacquey, Bruno Viaris de Lesegno and Laurence Pruvost[1]
[1]Laboratoire Aimé Cotton, CNRS, Univ Paris-Sud, ENS Paris-Saclay, Univ Paris-Saclay,
bat 505, 91405 Orsay, France
e-mail address: laurence.pruvost@u-psud.fr

The orbital angular momentum (OAM) of light is a quantity explored for communication and quantum technology, its key strength being a wide set of values offering a large basis for q-working. In this context we have studied the vortex conversion from a red optical vortex to a blue one, for OAMs ranging $-30$ to $+30$. The conversion is realized in a rubidium vapor, via the $5S_{1/2} - 5D_{5/2}$ $^{85}$Rb two-photon transition done with a Gaussian beam at $780\ nm$ plus a Laguerre-Gaussian beam at $776\ nm$ with the OAM $\ell$, producing a radiation at $420\ nm$. With co-propagating input beams, we demonstrate a conversion from red to blue for high-$\ell$ input vortices. We show that the output blue vortex respects the azimuthal phase matching, has a size determined by the product of the input beams and a power decreasing with $\ell$ in agreement with their overlap. Its propagation indicates that the generated blue wave is a nearly pure Laguerre-Gaussian mode. The vortex converter thus permits a correct OAM transmission.

## I. INTRODUCTION

Optical vortices, because of their quantized orbital angular momenta [1], are now widely explored for quantum optics, communication and technologies [2]. Beside the polarization – a well-known quantum variable of light with two possible values – the orbital angular momentum (OAM) theoretically ranges the $[-\infty,+\infty]$ interval and in principle offers the possibility to work with a huge space of quantum variables. In the optical domain the state of the art of OAM generation is $\sim 100\hbar$ for reconfigurable liquid crystals devices and $\sim 10\ 000\hbar$ for non-reconfigurable spiral phase mirrors [3]. Such values open possibilities for classical multiplexing [4] and quantum applications.

Non-linear processes with OAMs have been experienced in atomic vapors, cold or not, for demonstrating storage and retrieval [5], amplification [6] or OAM transfer [7]. Among these processes, frequency-changing ones, e.g. from red to blue as in Ref. [7], offer the great advantage of an unambiguous output signal. These experiments have been successfully demonstrated for small OAM values. To overcome the challenge of OAM use in a wide quantum basis, it is thus fundamental to explore these processes and characterize them with respect to large OAM values.

In this paper we report on vortex-conversion realized for OAMs ranging $[-30,+30]$ and associated to a frequency change from $776\ nm$ to $420\ nm$. Figure 1 illustrates the principle. The conversion uses the rubidium $5S_{1/2}$-$5D_{5/2}$ two-photon transition [8], realized with two co-propagating lasers, respectively at $\lambda_1 = 780\ nm$ and $\lambda_2 = 776\ nm$, the laser at $\lambda_2$ being a Laguerre-Gaussian (LG) mode with an azimuthal number $\ell_2 = \ell$ (its OAM) whereas the laser at $\lambda_1$ is a Gaussian (G) one, so $\ell_1 = 0$. Both lasers are detuned from the $5P_{3/2}$ intermediate level, to minimize its population, its decay to the $5S_{1/2}$ level and to optimize the population of $5D_{5/2}$ level [9]. Then, the stimulated coherent emission generates a photon pair, respectively at $\lambda_{IR} = 5.23\ \mu m$ and $\lambda_b = 420\ nm$. The process is known as four wave mixing (FWM) induced by amplified spontaneous emission (ASE).

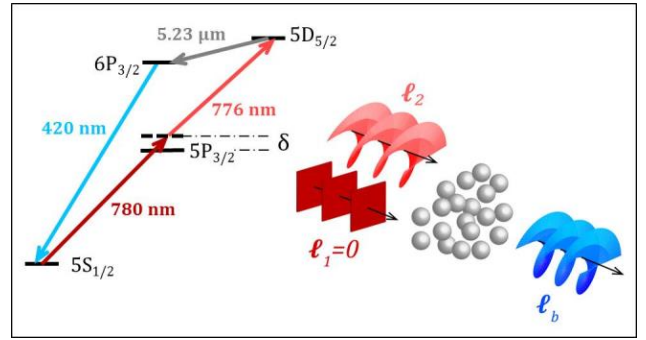

FIG. 1. Principle of vortex conversion based on the two-photon transition in Rubidium (scheme on the left).

With non-focused coaxial input beams, we show that the blue output vortex respects $\ell_b = \ell$ and that the conversion is efficient up to $|\ell| = 30$. This differs from previous experiments realized with focused input beams [7] that convert small OAMs. We also show that the blue output vortex has a ring size determined by the product of the input beams, a power decreasing with $|\ell|$ with respect to the input beams overlap and a propagation attesting a high-quality mode with a M$^2$ factor close to 1.

## II. THE EXPERIMENT

The experimental scheme is shown in Fig. 2. The vortex conversion is realized in a 10-cm long, 25 mm-diameter rubidium vapor cell which is heated and maintained at 130°C. The atomic density is then about $5 \times 10^{13}$ at/cm$^3$.

The atoms are illuminated by two co-propagating laser beams, one at $\lambda_1 = 780\ nm$ being a Gaussian beam, the second at $\lambda_2 = 776\ nm$ being a $\ell$-vortex (a LG mode with the OAM $\ell$). The two lasers are superimposed and circularly polarized with same handedness for efficiently realizing the $5S_{1/2}$-$5D_{5/2}$ two-photon transition [9].

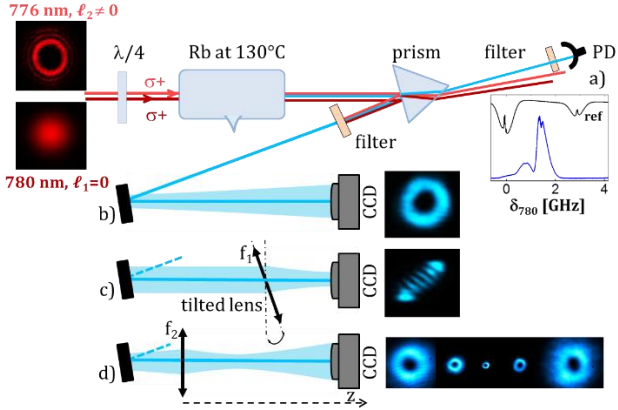

FIG. 2. Experimental setup : a) PD detection giving the blue power and the FWM spectrum (blue) as $\lambda_1$ is scanned and the 5s-5p saturated absorption spectrum (black) for calibration; b) beam profile detection; c) OAM detection by the tilted lens method; d) propagation detection.

The laser at $\lambda_1$ is provided by an amplified laser diode (Toptica TaPro not represented in Fig. 2), delivering a $1\ W$, $1\ MHz$-width radiation. Its frequency is controlled using a saturated absorption spectroscopy setup (not represented in Fig. 2) whose signal is shown by the black curve in Fig. 2(a). The $\lambda_1$-beam is fibered and routed to the cell. With a fiber-collimator and lenses the beam is size-arranged to a waist $w_1 = 0.17\ mm$ (so its Rayleigh range is $z_{R1} = 12\ cm$) located inside the cell. The light power $P_1$ illuminating the atoms is about 100 mW.

The $\ell$-vortex at $\lambda_2$ is provided by a Ti-Sa laser (Coherent MBR-110 not represented in Fig. 2) delivering 1W, 1 $MHz$ width radiation. Its frequency is controlled by a 30 $MHz$-resolution wavemeter (Burleigh WA-1100 not represented). The beam is transformed into a $\ell$-vortex by imprinting a helical phase of helicity $\ell$ on its wavefront. To do that, we use a spatial light modulator (SLM Hamamatsu PPM X8267 not represented) in phase-modulation configuration, addressed by a computed hologram. The generated $\ell$-vortex is close to a LG mode with a radial number $p = 0$ and an azimuthal number $\ell$. The $\ell$-vortex is then size-arranged to get a ring radius $R_\ell$ and a Rayleigh range not too far from $w_1$ and $z_{R1}$. As observed in [10[10]], $R_\ell$ is linear with $\ell$ (see Fig. 3(b)). It ranges from ~$0.07\ mm$ for $\ell = 1$ to ~$0.7\ mm$ for $\ell = 30$. The corresponding Rayleigh range [11[11]] varies from 4 cm for $\ell = 1$ to 14 cm for $\ell = 30$. A power of $\ell$-vortex beam $P_2$ of about $100\ mW$ is sent to the cell.

The blue emission relies on the population transfer from $5S_{1/2}$ level to $5D_{5/2}$ level, which is optimum for a laser detuning $\delta$ ranging 1-2 $GHz$ [9]. With our laser powers, we have found an efficient blue emission at $\delta \approx 1.5\ GHz$. To understand this value let us check at the Rabi frequencies related to the excitations. Note that the Rabi frequency $\Omega_1$ of the $\lambda_1$-excitation to be considered in the process is not the value at the beam center (about $1.6\ GHz$) but the value at the radius $R_\ell$ which is lower (typically $\Omega_1 \sim 0.1\ GHz$ for $R_{\ell=10}$) and guaranties $\Omega_1^2/\delta^2 \ll 1$ and so a weak $5P_{3/2}$ population. The value of $\Omega_2$ associated to the $\lambda_2$-excitation is obtained at the maximum intensity of the $\ell$-vortex, so located at $R_\ell$. We get $\Omega_2 \sim 0.2\ GHz$ for $\ell = 10$. Finally the effective Rabi frequency of the two-photon transition is $\Omega_1\Omega_2/(2\delta) \approx 10\ MHz$. Compared to $5D_{5/2}$ linewidth equal to $0.4\ MHz$, that guaranties a full population transfer to the $5D_{5/2}$ level.

In our experiment, because the IR light at $\lambda_{IR} = 5.23\ \mu m$ is absorbed by the glass cell, we only detect and analyze the blue output beam. For that, we separate it from the red input beams either by a prism or bandpass filters. Then, the blue light is analyzed with four detection branches, which are denoted from a) to d) in Fig. 2.

Detection a) is a photodiode (PD) which records the spectral response of the FWM process (blue line of Fig 2(a)) as $\lambda_1$ is scanned. Then, we fix $\lambda_1$ at the maximum of the emission and measure the blue power via the PD voltage. Typically, we get 80 $\mu W$ for $\ell = 1$ and 120 $nW$ for $\ell = 30$.

Detection b) is a CCD camera placed at $z_b = 400\ mm$ far from the cell which takes a picture of the created blue vortex. It allows us to determine its profile and its radius.

Detection c) is a CCD camera placed close to the focus point of a lens $f_1 = 500\ mm$ to take a picture of the created blue vortex. As the lens $f_1$ is tilted (typically by ~25°), this astigmatic device creates an auto-interference of the vortex. The number of dark fringes of the pattern being equal to the azimuthal number [12[12]] allows us to determine the blue vortex OAM.

Detection d) is a CCD camera moving on a translation rail behind a lens $f_2 = 200\ mm$ to take pictures of the blue beam at different $z$ positions and to analyze its propagation.

## III. CONVERSION VS THE INPUT OAM

As first study, we have analyzed the blue intensity profile and its OAM versus $\ell$. The profiles recorded by detection b) show that, for any $\ell$ value, the blue beam exhibits a ring profile (see examples in Fig. 3). The measured blue radius $R_b$ plotted versus $\ell$ is shown in Fig.3(b) to be compared to the input one $R_\ell$. Although the input radius $R_\ell$ variation is linear, the output $R_b$ is not.

With detection c) we have checked the OAM of the beams. As illustrated in Fig. 3(a), the number of dark fringes are the same for the red input and the blue output. So we deduce $\ell_b = \ell$. This conclusion is without ambiguity for $|\ell| \leq 11$. For higher values, with our method, it was difficult to distinctly count the number of dark fringes; the error is estimated to 10%. As a conclusion, our measurement corroborates that mainly $\ell_{IR} = 0$, as observed in [7], and that the transfer occurs from the beam at 776 nm to the blue output.

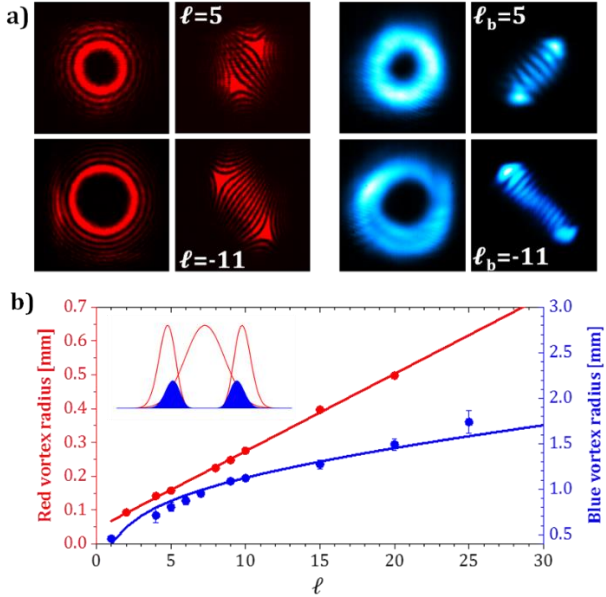

FIG. 3. Blue output vs $\ell$-vortex input. a) Intensities profiles and corresponding auto-interference patterns, input on the left; output on the right. b) Ring radii vs $\ell$. Input $R_\ell$ (red dots) measured at the cell, fitted by $R_\ell = R_0(1+\beta\ell)$ with $R_0 = 0.045 \pm 0.004\ mm$, $\beta = 0.51 \pm 0.07$. Output $R_b$ (blue dots) measured at $z_b$ from the cell, fitted by Eq. (2) corrected by $\sqrt{1+z_b^2/z_{Rb}^2}$ where $z_{Rb} = \pi w_{12}^2/\lambda_b$ with $w_1 = 0.15 \pm 0.02\ mm$, $R_0 = 0.06 \pm 0.01\ mm$, $\beta = 0.51$ (fixed). The inset shows $I_1, I_2$ (red lines) and the product (blue line).

The two observations, i.e. $\ell_b = \ell$ and $R_b$ nonlinear versus $\ell$, are explained by considering the FWM process in the weak gain regime. Indeed, the FWM process which produces the blue field $E_b$ is described by a nonlinear term $\chi^3 E_1 E_2 E_{IR}^*$, where $\chi^3$ is the non-linear susceptibility and $E_i$ ($i = 1, 2, IR$) are the involved fields. As a consequence, the generation of $E_b$ is realized under phase-matching conditions, one of them being OAM conservation. In addition, in the weak gain regime, the amplitude $E_b$ is known to be proportional to the input field product $E_1 E_2$ whose shape will explain the observed output radius $R_b$.

Let us give precisions on phase-matching. It concerns the propagation phase, the azimuthal phase, the curvature phase and the Gouy phase. In the collinear beam geometry as used here, because of energy conservation, the propagation phase is conserved [13] giving $k_b + k_{IR} = k_1 + k_2$ where $k_i$ are the wave vectors. The azimuthal phase-matching [14], for collinear beams, implies $\ell_b + \ell_{IR} = \ell_1 + \ell_2$. With $\ell_1 = 0$ and $\ell_2 = \ell$ a priori many pairs $\{\ell_b, \ell_{IR}\}$ with $\ell_b + \ell_{IR} = \ell$ are possible. But, because IR light is mainly created by ASE, then $\ell_{IR} = 0$ is mainly expected. In addition because $\lambda_{IR}$ and $\lambda_b$ are very different, to get Rayleigh ranges and Gouy phases of same order of magnitude, $\ell_{IR} = 0$ is the most favorable case. It explains why $\ell_b = \ell$ is experimentally observed.

The curvature and the Gouy phases are radially and longitudinally dependent [15]. The curvatures being large, the curvature phases are nearly zero and correctly matched. For the Gouy phase, the matching leads to [16]

$$\frac{\lambda_b(|\ell_b|+1)}{w_b^2} + \frac{\lambda_{IR}(|\ell_{IR}|+1)}{w_{IR}^2} = \frac{\lambda_1}{w_1^2} + \frac{\lambda_2(|\ell|+1)}{w_2^2} \quad (1)$$

Note that, if the Rayleigh ranges are equal (Boyd criterion [17]), then Eq. (1) gives $|\ell_b| + |\ell_{IR}| = |\ell|$ which is obviously satisfied if $\ell_{IR} = 0$ and $\ell_b = \ell$. Else, with $\lambda_1 \approx \lambda_2$ and $\lambda_{IR} \gg \lambda_b, \lambda_1$ the analysis of the solutions of Eq. (1) shows that in many cases, $w_{IR} \gg w_1$ and $\ell_{IR} = 0$ is the most probable solution.

Let us now consider the weak gain regime. For strong input fields $E_1$, $E_2$, the output blue field amplitude is proportional to the product of the inputs, so $\sqrt{I_b} \propto Z\sqrt{I_1 I_2}$ where $I_1$ and $I_2$ are the beam intensities and $Z$ is the amplification length. In practice, $Z$ is the cell length for beams having long Rayleigh ranges, else $Z$ is shorter. As a consequence, the blue intensity is proportional to $I_1 I_2$ which thus determines (i) the ring size of the blue vortex and (ii) the conversion efficiency (see the following section).

For input beams, being a G and a one-ring LG mode, $I_1$ and $I_2$ are written as

$$I_1 = \frac{2P_1}{\pi w_1^2} e^{\frac{-2r^2}{w_1^2}}\ ;\ I_2 = \frac{2P_2}{\pi w_2^2 \ell!}\left(\frac{2r^2}{w_2^2}\right)^{|\ell|} e^{\frac{-2r^2}{w_2^2}}$$

Then, the product $I_1 I_2$ has a ring shape (see inset of Fig.3b) whose radius is $R_{12} = w_{12}\sqrt{|\ell|/2}$ where $w_{12} = w_1 w_2/\sqrt{w_1^2 + w_2^2}$. Subsequently we obtain

$$R_{12} = R_\ell \frac{1}{\sqrt{1 + 2R_\ell^2/(|\ell|w_1^2)}} \quad (2)$$

Note that $R_{12}$ is not proportional to $R_\ell$, so no longer linear with $\ell$. Eq. (3) has been used to fit the data in Fig. 3(b). We have just added the propagation correction over $z_b = 400\ mm$. Model and data are in a good agreement.

## IV. CONVERSION EFFICIENCY.

In a second study we have examined the conversion efficiency. For that, the blue power $P_b$ measured with the detection a) is compared to the red input power at $776\ nm$ measured at the cell entrance. We have deduced the vortex-conversion efficiency (the ratio). Figure 4 which spans the data for $\ell$ ranging $[-30, +30]$, in logarithmic scale, shows an efficiency which rapidly drops down as $|\ell|$ increases, typically one decade from $|\ell|=1$ to 10. Data slightly depend on the sign of $\ell$ (difference~10%), probably due to a setup asymmetry in the $\ell$-vortex generation which produces radii slightly larger for $\ell<0$.

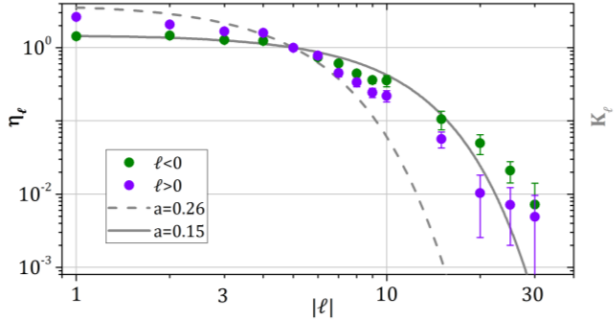

FIG. 4. Vortex-conversion efficiency: experimental data $\eta_\ell = P_b/P_2$ (dots), overlap $K_\ell$, for $a = R_0/w_1 = 0.15$ (line) and $0.26$ (dashed line) and $\beta = 0.51$. All curves have been normalized to the case $|\ell| = 5$ [18[18]].

In the weak gain regime, because the intensity $I_b$ is proportional to $I_1 I_2$ we deduce that blue power is proportional to the quantity $K_\ell = \int_0^\infty I_1(r) I_2(r) 2\pi r dr$. This quantity evaluated for a G and a LG mode is $K_\ell = \left(\frac{2P_1 P_2}{\pi w_1^2}\right)/\left(1 + \frac{w_2^2}{w_1^2}\right)^{|\ell|+1}$ By introducing in this formula, the input $w_2 = R_0(1 + \beta|\ell|)/\sqrt{|\ell|/2}$ we get the curves shown in Fig. 4. The agreement with the experiment obtained for $a = R_0/w_1 = 0.15$ allows us to conclude that vortex-conversion is routed by the field overlap. Then it could be optimized by adjusting the Gaussian beam waist.

## V. BLUE VORTEX PROPAGATION

In the last study we have analyzed the blue beam propagation to investigate the mode purity. Using detection d) we have recorded the blue wave shape along the propagation, measured the horizontal and vertical radii and plotted the variation versus $z$ [see data for $\ell$=4 and 8 plot in Fig. 5].

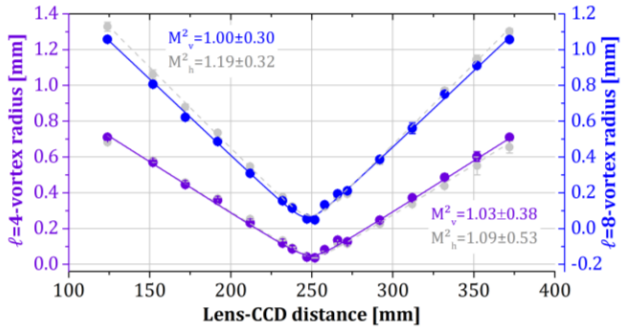

FIG. 5. Propagation of the blue vortex: vertical radii (dots), horizontal radii (gray dots) for $\ell = 4$ (violet) and $\ell = 8$ (blue). Fits (lines) giving the $M^2$ factors (indicated close to the plots).

Propagations curves have been fitted by the hyperbola function $w_0\sqrt{|\ell|/2}\sqrt{1 + (z - z_0)^2/z_R^2}$ where $w_0$ is the waist at $z = z_0$ and $z_R$ the Rayleigh range. It provides the $M^2$ factor, as $M^2 = (\pi w_0^2/\lambda_b)/z_R$. We find $M^2$ very close to 1 for both horizontal and vertical analysis, for $\ell = 4$ and $\ell = 8$ (see Fig. 5) which attests the purity of the blue mode.

The good purity of the created vortex, compared to the input $\ell$-vortex one (see Fig. 3a [19]), is explained by the non-linear source term. Being a field product this terms acts as a mode cleaner especially when $R_\ell$ is larger than $w_1$. The quality of the blue vortex is a great advantage for carrying information over a long distance or for using it as a pure LG source.

## VI. CONCLUSION

The vortex-conversion performed in a Rb vapor in a nearly parallel geometry has shown an OAM transfer for high helicity waves [-30,+30], to a wave with a different color. The efficiency is mainly governed by the input field overlap. Furthermore we show that the created blue vortex is LG mode with a high purity. That guaranties a good fidelity of the device viewed as an OAM transmitter and opens possibilities in the field of quantum variable manipulation (here the OAM) for quantum communication.

## ACKNOWLEDGEMENTS

The experiment has been financed by CNRS, Université Paris-Sud (BQR-grant), Université Paris-Saclay (labex PALM) and Région Ile de France (DIM nanoK). The authors thank L. Bizet and B. Jost who participated during their internships and C. Drag from LPP (Laboratoire de Physique des Plasmas of Palaiseau) for stimulating discussions.
[2] L. Allen, M. W. Beijersbergen, R. J. C. Spreeuw and J. P. Woerdman, Phys. Rev. A **45**, 8185 (1992).
[1] For a review : M. J. Padgett, Optics. Expr. 25, 11266 (2017) and the references therein; M. Barnett, M. Babiker and M. J. Padgett, Phil. Trans. R. Soc. A **375,** 2087 (2017).
[3] R. Fickler, G. T. Campbell , B. C. Buchler , P. K. Lam , A. Zeilinger , Proceedings of the National Academy of Sciences **113** (48), 13642 (2016).
[4] J. Wang, J. Yang, I. M. Fazal, N. Ahmed, Y. Yan, H. Huang, Y. Ren, Y. Yue, S. Dolinar, M. Tur and A. E. Willner, Nature phot. **6**, 488 (2012); R. Fickler, R. Lapkiewicz, W. N. Plick, M. Krenn, C. Schaeff, S. Ramelow, A. Zeilinger, Science 338, 640 (2012); F. Tamburini, E. Mari, A. Sponselli, B. Thidé, A. Bianchini, F. Romanato, New J. Phys. **14**, 033001 (2012); M. Krenn, R. Fickler, M. Fink, J. Handsteiner, M. Malik, T. Scheidl, R. Ursin, A. Zeilinger, New J. P. **16**, 113028 (2014).

[5] J. W. R. Tabosa and D. V. Petrov, Phys. Rev. Lett. **83**, 4967 (1999); W. Jiang, Q.-F. Chen, Y.-S. Zhang, and G.-C. Guo, Phys. Rev. A **74**, 043811 (2006); R. Pugatch, M. Shuker, O. Firstenberg, A. Ron, and N. Davidson, Phys. Rev. Lett. **98**, 203601 (2007); A. Nicolas, L. Veissier, L. Giner, E. Giacobino, D. Maxein and J. Laurat, Nat. photonics **8**, 234 (2014); A. J. F. de Almeida, S. Barreiro, W. S. Martins, R. A. de Oliveira, D. Felinto, L. Pruvost, and J. W. R. Tabosa, Opt. Lett. **40**, 2545 (2015); D.-S. Ding, W. Zhang, Z.-Y. Zhou, Shuai Shi, G.-Y. Xiang, X.-S. Wang, Y.-K. Jiang, B.-S. Shi, and G.-C. Guo, Phys. Rev. Lett. **114**, 050502 (2015).

[6] G. C. Borba, S. Barreiro, L. Pruvost, D. Felinto and J. W. R. Tabosa, Opt. Exp. **24**, 10078 (2016).

[7] G. Walker, A. S. Arnold, and S. Franke-Arnold, Phys. Rev. Lett. **108**, 243601 (2012); A. M. Akulshin, R. J. McLean, E. E. Mikhailov, and I. Novikova, Opt. Lett. **40**, 1109 (2015).

[8] F. Nez, F. Biraben, R. Felder, Y. Millerioux, Optics Comm. **102**, 432 (1993); T. T. Grove, V. Sanchez-Villicana, B.C. Duncan, S.Maleki, P. L. Gould, Physica Scripta **52**, 271 (1995).

[9] A. S. Zibrov, M. D. Lukin, L. Hollberg, M.O. Scully, Phys. Rev. A **65**, 051801(R) (2002); T. Meijer, J. D. White, B. Smeets, M. Jeppesen, R. E. Scholten, Opt. Lett. **31,** 1002 (2006); A. M. Akulshin, R. J. McLean, A. I. Sidorov and P. Hannaford, Opt. Exp. **17**, 22861 (2009); A. Vernier, S. Franke-Arnold, E. Riis and A.S. Arnold, Opt. Exp. **18**, 17020 (2010); J. F. Sell, M. A. Gearba, B. D. DePaola and R. J. Knize, Opt. Lett. **39**, 3, 528 (2014); A. Akulshin, D. Budker and R. McLean, Opt. Lett. **39**, 845 (2014).

[10] J. E. Curtis and D. G. Grier, Phys. Rev. Lett **90**, 133901, (2003).

[11] For a one-ring LG mode (radial number equal to zero), radius and waist are related by $R_\ell = w_2\sqrt{|\ell|/2}$; the Rayleigh range being $z_{R2} = \pi w_2^2/\lambda_2$ is then $z_{R2} = 2\pi R_\ell^2/(\lambda_2|\ell|)$.

[12] P. Vaity, J. Banerji, R. P. Singh, Phys. Lett. A **377**, 1154 (2013); N. C. Cabrera Gutiérrez, Thèse de doctorat de l'Université Paris Sud, 2014 NNT : 2014PA112360. https://tel.archives-ouvertes.fr/tel-01216481v2.

[13] The vapor is diluted, so we consider the refractive indices close to 1. The propagation phase is $k_i z$.

[14] The azimuthal phase is $\ell_i \theta$, where $\theta$ is the polar angle in the cylindrical coordinates oriented by $z$.

[15] For a one-ring LG mode the curvature phase is $k_i r^2/2\mathcal{R}(z)$ where $\mathcal{R}(z) = z(1 + z_R^2/z^2)$. The Gouy phase is $(|\ell_i| + 1)Atan(z/z_{Ri})$.

[16] The conservation of the Gouy phase implies $\sum \eta_i (|\ell_i| + 1)Atan(z/z_{Ri}) = 0$ where $\eta_i = \pm 1$ for input/output waves. The derivative versus $z$, taken at $z = 0$, leads to $\sum \eta_i (|\ell_i| + 1)/z_{Ri} = 0$.

[17] G. D. Boyd and D. A. Kleinman, J. Appl. Phys. **39**, 3597 (1968).

[18] As soon as $|\ell| \geq 5$ the population of the $5p$ level is negligible and the population of 5d level is maximum.

[19] Note the outer rings for the input $\ell$-vortex (Fig.3(a)) indicating that the mode is not perfectly pure.

[1] Orbital angular momentum of light and the transformation of Laguerre-Gaussian laser modes, L. Allen, M. W. Beijersbergen, R. J. C. Spreeuw and J. P. Woerdman, Phys. Rev. A **45**, 8185 (1992).

[2] For a review: Orbital angular momentum 25 years on, M. J. Padgett, Optics. Expr. 25, 11266 (2017) and the references therein; Optical orbital angular momentum, by S. M. Barnett, M. Babiker and M. J. Padgett, Phil. Trans. R. Soc. A **375**, 2087 (2017).

[3] Quantum entanglement of angular momentum states with quantum numbers up to 10010, R. Fickler, G. T. Campbell, B. C. Buchler, P. K. Lam, A. Zeilinger, Proceedings of the National Academy of Sciences **113** (48), 13642 (2016).

[4] Terabit free space data transmission employing orbital angular momentum multiplexing, J. Wang, J. Yang, I. M. Fazal, N. Ahmed, Y. Yan, H. Huang, Y. Ren, Y. Yue, S. Dolinar, M. Tur and A. E. Willner, Nature phot. **6**, 488 (2012).

Quantum entanglement of high angular momenta, R. Fickler, R. Lapkiewicz, W. N. Plick, M. Krenn, C. Schaeff, S. Ramelow, A. Zeilinger, Science **338**, 640 (2012)

Encoding many channels on the same frequency through radio vorticity: first experimental test, F. Tamburini, E. Mari, A. Sponselli, B. Thidé, A. Bianchini, F. Romanato, New J. Phys. **14**, 033001 (2012).

Communication with spatially modulated light through turbulent air across Vienna, M. Krenn, R. Fickler, M. Fink, J. Handsteiner, M. Malik, T. Scheidl, R. Ursin, A. Zeilinger, New J. P. **16**, 113028 (2014).

[5] Optical Pumping of Orbital Angular Momentum of Light in Cold Cesium Atoms J. W. R. Tabosa and D. V. Petrov, Phys. Rev. Lett. **83**, 4967 (1999).

Computation of topological charges of optical vortices via nondegenerate four-wave mixing, W. Jiang, Q.-F. Chen, Y.-S. Zhang, and G.-C. Guo, Phys. Rev. A **74**, 043811 (2006).

Topological Stability of stored Optical Vortices, R. Pugatch, M. Shuker, O. Firstenberg, A. Ron, and N. Davidson, Phys. Rev. Lett. **98**, 203601 (2007).

A quantum memory for orbital angular momentum photonic qubits A. Nicolas, L. Veissier, L. Giner, E. Giacobino, D. Maxein and J. Laurat, Nat. photonics **8**, 234 (2014).

Storage of orbital angular momenta of light via coherent population oscillation, A. J. F. de Almeida, S. Barreiro, W. S. Martins, R. A. de Oliveira, D. Felinto, L. Pruvost, and J. W. R. Tabosa, Opt. Lett. **40**, 2545 (2015).

Quantum Storage of Orbital Angular Momentum Entanglement in an Atomic Ensemble, D.-S. Ding, W. Zhang, Z.-Y. Zhou, Shuai Shi, G.-Y. Xiang, X.-S. Wang, Y.-K. Jiang, B.-S. Shi, and G.-C. Guo, Phys. Rev. Lett. **114**, 050502 (2015).

[6] Narrow band amplification of light carrying orbital angular momentum, G. C. Borba, S. Barreiro, L. Pruvost, D. Felinto and J. W. R. Tabosa, Opt. Exp. **24**, 10078 (2016).

[7] Trans-Spectral Orbital Angular Momentum Transfer via Four-Wave Mixing in Rb Vapor, G. Walker, A. S. Arnold, and S. Franke-Arnold, Phys. Rev. Lett. **108**, 243601 (2012).

Distinguishing nonlinear processes in atomic media via orbital angular momentum transfer, A. M. Akulshin, R. J. McLean, E. E. Mikhailov, and I. Novikova, Opt. Lett. **40**, 1109 (2015).

[8] Optical frequency determination of the hyperfine components of the 5S1/2-5D3/2 two-photon transitions in rubidium, F. Nez, F. Biraben, R. Felder, Y. Millerioux, Optics Comm. **102**, 432 (1993).

Two photon two-color diode laser spectroscopy of Rb 5D5/2 state, T. T. Grove, V. Sanchez-Villicana, B.C. Duncan, S.Maleki, P. L. Gould, Physica Scripta **52**, 271 (1995).

[9] Efficient frequency up-conversion in resonant coherent media, A. S. Zibrov, M. D. Lukin, L. Hollberg, M.O. Scully, Phys. Rev. A **65**, 051801(R) (2002)

Blue five-level frequency-upconversion system in rubidium, T. Meijer, J. D. White, B. Smeets, M. Jeppesen, R. E. Scholten, Opt. Lett. **31,** 1002 (2006)

Coherent and collimated blue light generated by four wave mixing in Rb vapor, A. M. Akulshin, R. J. McLean, A. I. Sidorov and P. Hannaford, Opt. Exp. **17**, 22861 (2009)

Enhanced frequency up-conversion in Rb vapor, A.Vernier, S. Franke-Arnold, E. Riis and A.S. Arnold, Opt. Exp. **18**, 17020 (2010),

Collimated blue and infrared beams generated by two photons excitation in Rb vapor, J. F. Sell, M. A. Gearba, B. D. DePaola and R. J. Knize, Opt. Lett. **39**, 3, 528 (2014)

Directional infrared emission resulting from cascade population inversion and four wave mixing in Rb vapours, A. Akulshin, D. Budker and R. McLean, Opt. Lett. **39**, 845 (2014).

[10] Structure of Optical Vortices, J. E. Curtis and D. G. Grier, Phys. Rev. Lett **90**, 133901 (2003).

[11] For a one-ring LG mode (radial number equal to zero), radius and waist are in relation by $R_\ell = w_2\sqrt{|\ell|/2}$; the Rayleigh range is $z_{R2} = \pi w_2^2/\lambda_2$, then $z_{R2} = 2\pi R_\ell^2/(\lambda_2|\ell|)$.

[12] Measuring the topological charge of an optical vortex by using a tilted convex lens, P. Vaity, J. Banerji, R. P. Singh, Phys. Lett. A **377**, 1154 (2013).

Modes de Laguerre-Gauss et canalisation d'atomes froids, N. C. Cabrera Gutiérrez, Thèse de doctorat de l'Université Paris Sud - 2014 NNT : 2014PA112360. https://tel.archives-ouvertes.fr/tel-01216481v2

[13] The vapor is diluted, so we consider the refractive indices close to 1. The propagation phase is $k_i z$.

[14] The azimuthal phase is $\ell_i \theta$, where $\theta$ is the polar angle in the cylindrical coordinates oriented by $z$.

[15] For a one-ring LG mode the curvature phase is $k_i r^2/2\mathcal{R}(z)$ where $\mathcal{R}(z) = z(1 + z_R^2/z^2)$. The Gouy phase is $(|\ell_i| + 1)Atan(z/z_{Ri})$.

[16] The conservation of the Gouy phase implies $\sum \eta_i (|\ell_i| + 1)Atan(z/z_{Ri}) = 0$ where $\eta_i = \pm 1$ for input/output waves. The derivative versus $z$, taken at $z = 0$, leads to $\sum \eta_i \ (|\ell_i| + 1)/z_{Ri} = 0$.

[17] G. D. Boyd and D. A. Kleinman, J. Appl. Phys. **39**, 3597 (1968).

[18] As soon as $|\ell| \geq 5$ the population of the $5p$ level is negligible and the population of 5d level is maximum.

[19] Note the outer rings for the input $\ell$-vortex (Fig.3(a)) indicating that the mode is not perfectly pure.